%% file: paper.tex
\documentclass{aa}

\usepackage[varg]{txfonts}

\input{definitions.tex}

\begin{document}

\title{Detection of infrared fluorescence of carbon dioxide in R~Leonis with SOFIA/EXES}
\titlerunning{Infrared fluorescence of CO$_2$ in \rleo}
\authorrunning{J. P. Fonfr\'{\i}a et al.}
\author{J.~P.~Fonfr\'ia\inst{\ref{csic}}
  \and E.~J.~Montiel\inst{\ref{usra}}
  \and J.~Cernicharo\inst{\ref{csic}}
  \and C.~N.~DeWitt\inst{\ref{usra}}
  \and M.~J.~Richter\inst{\ref{ucdavis}}}
\institute{
  Molecular Astrophysics Group, Instituto de F\'isica Fundamental, IFF-CSIC, C/ Serrano, 123, 28006, Madrid (Spain);~\email{jpablo.fonfria@csic.es}\label{csic}
  \and
  SOFIA-USRA, NASA Ames Research Center, MS 232-12, Moffett Field, CA 94035 (USA)\label{usra}
  \and
  Physics Dept. - UC Davis, One Shields Ave., Davis, CA 95616 (USA)\label{ucdavis}
}

\abstract{
We report on the detection of hot CO$_2$ in the O-rich AGB star \rleo{} based on high spectral resolution observations in the range $12.8-14.3$~$\mu$m carried out with the Echelon-cross-Echelle Spectrograph (EXES) mounted on the Stratospheric Observatory for Infrared Astronomy (SOFIA).
We have found $\simeq 240$ CO$_2$ emission lines in several vibrational bands.
These detections were possible thanks to a favorable Doppler shift that allowed us to avoid contamination by telluric CO$_2$ features.
The highest excitation lines involve levels at an energy of $\simeq 7000$~K.
The detected lines are narrow (average deconvolved width~$\simeq 2.5$~\kms) and weak (usually $\lesssim 10\%$ the continuum).
A ro-vibrational diagram shows that there are three different populations, warm, hot, and very hot, with rotational temperatures of $\simeq 550$, 1150, and 1600~K, respectively.
From this diagram, we derive a lower limit for the column density of $\simeq 2.2\times 10^{16}$~\cmm.
Further calculations based on a model of the \rleo{} envelope suggest that the total column density can be as large as $7\times 10^{17}$~\cmm{} and the abundance with respect to H$_2$ $\simeq 2.5\times 10^{-5}$.
The detected lines are probably formed due to de-excitation of CO$_2$ molecules from high energy vibrational states, which are essentially populated by the strong \rleo{} continuum at 2.7 and 4.2~$\mu$m.
}

\maketitle

\section{Introduction}
\label{sec:introduction}

Extensive work has been done during the last decades to characterize the molecular content of the circumstellar envelopes (CSEs) of asymptotic giant branch stars (AGBs).
This huge effort has been fruitful having found more than 90 different molecular species, not counting isotopologues \citep{mcguire_2018}, most of which have been detected for first time in the carbon-rich star \irc{} \citep[e.g.,][]{ridgway_1976,betz_1981,goldhaber_1984,bernath_1989,cernicharo_2000,agundez_2008,agundez_2014}.
Oxygen-rich (O-rich) stars are less chemically active than carbon stars but many molecules have also been observed in their envelopes.
Among them, we can find ubiquitous species such as CO, HCN, SiS, SiO or CS and other molecules typically formed in O-rich environments such as H$_2$O, SO, SO$_2$, H$_2$CO or NO \citep[e.g.,][]{velilla-prieto_2017}.

Most of these molecules have been found in the millimeter range, where molecules without a permanent dipole moment cannot be observed due to their lack of a pure rotational spectrum.
One of them is CO$_2$, predicted to be an abundant parent species by chemical models \citep[e.g.,][]{cherchneff_2006,agundez_2020} with abundances related to H$_2$ of $10^{-8}-10^{-4}$.
The detections in AGB stars carried out to date were done only from space with the Infrared Space Observatory (\textit{ISO}) and the Spitzer Space Telescope \citep[e.g.,][]{justtanont_1996,justtanont_1998,ryde_1997,tsuji_1997,yamamura_1999,cami_2000,markwick_2000,sloan_2010,smolders_2012,reiter_2015,baylis-aguirre_2020} as a consequence of the extremely high atmospheric opacity.

Chemical models indicate that a large fraction of the initially available oxygen atoms are locked into CO$_2$ through the two-body chemical reactions of CO with H$_2$O and OH or the three-body reaction of CO and atomic oxygen assisted by a catalyst, efficient in high density environments.
Consequently, the formation efficiency of other O-bearing species is hampered by the formation of CO$_2$.
Despite this, CO$_2$ emission has been poorly analyzed in detail.

\rleo{} is one of the closest O-rich stars with a distance of $70-85$~pc and a low mass-loss rate of $\simeq 1.0\times 10^{-7}$~\mlr.
The terminal velocity of its expanding gas is $\simeq 6-9$~\kms{} \citep{debeck_2010,ramstedt_2014}.
The central star pulsates with a period of $\simeq 310$~days, its effective temperature is $\simeq 2500-3000$~K, and its angular diameter is $\simeq 0\farcs025-0\farcs030$ \citep[e.g.,][]{perrin_1999,fedele_2005,wittkowski_2016}.
The molecular observations carried out in the millimeter, submillimeter and infrared ranges \citep[e.g.,][]{hinkle_1979,bujarrabal_1994,etoka_1997,bieging_2000,gonzalez-delgado_2003,ohnaka_2004,schoier_2013} indicate that CO, H$_2$O, OH, SiO, SO, SO$_2$, and HCN exist with significant abundances, but CO$_2$ has not been detected toward \rleo{} so far.

In this Letter, we report on the detection of hot CO$_2$ toward the O-rich star \rleo{} performed from the Stratospheric Observatory for Infrared Astronomy \citep[SOFIA;][]{temi_2018} with the high spectral resolution Echelon-cross-Echelle Spectrograph \citep[EXES;][]{richter_2018}.

\begin{figure*}
  \centering
  \includegraphics[width=0.98\textwidth]{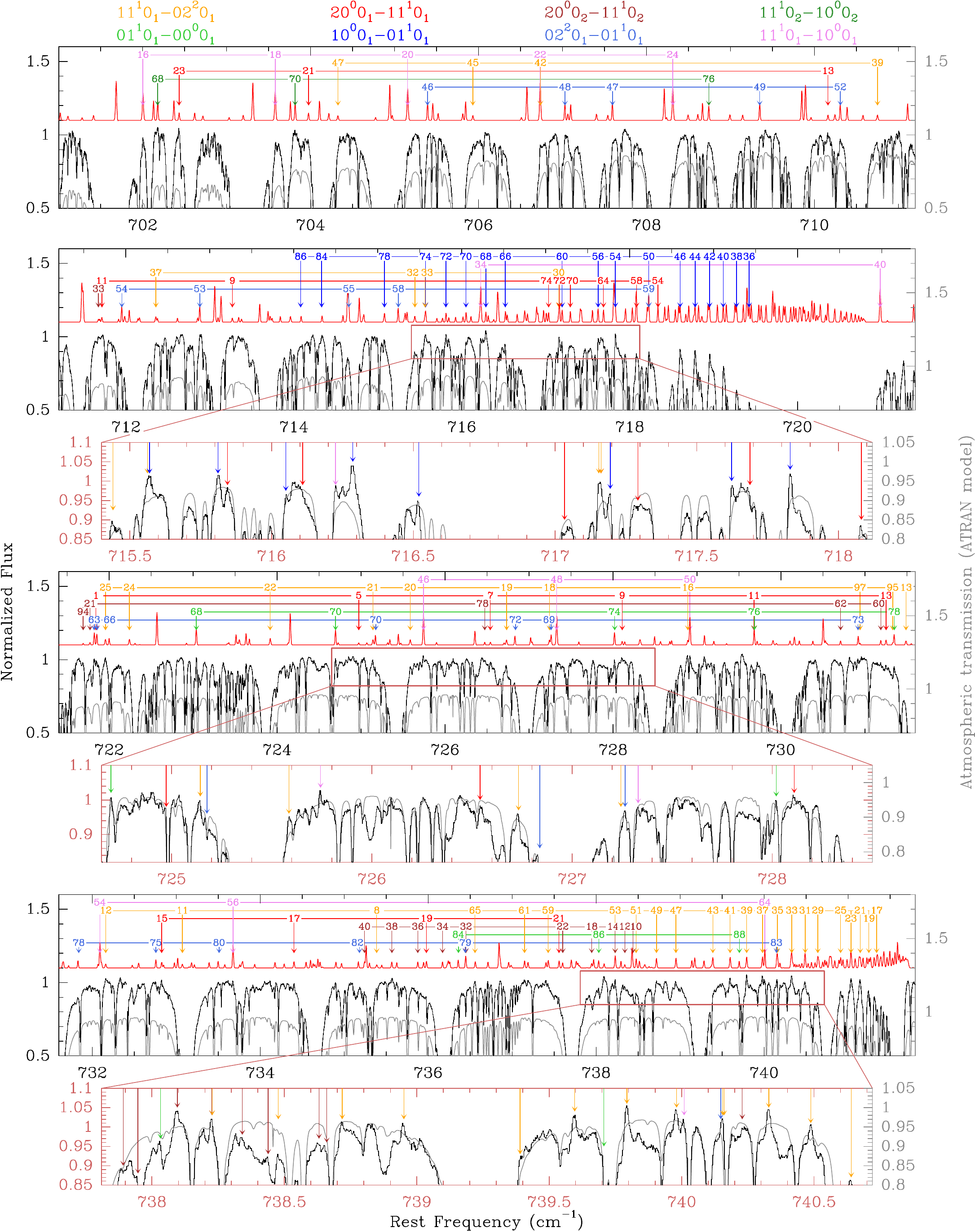}
  \caption{Spectrum of \rleo{} in the spectral range $701-742$~\cm{} in the rest frequency (i.e., we have corrected for \rleo's radial velocity with respect to Earth at the time of observation).
    The detected CO$_2$ are plotted in different colors depending on the band.
    The red synthetic spectrum is a model to the CO$_2$ emission (Sect.~\ref{sec:discussion}).
    The gray spectrum is the atmospheric transmission calculated with the ATRAN code \citep{lord_1992}.
    Only the lines clearly detected have been marked with arrows.
    The branches are not indicated for the sake of clarity.
    An additional spectral range is shown in Fig.~\ref{fig:f2}.}
  \label{fig:f1}
\end{figure*}


\section{Observations}
\label{sec:observations}

Observations of \rleo{} were carried out with SOFIA/EXES on 01 Nov 2018 (UT).
Two settings in the High--Low instrument configuration were taken, which yield a full range of 701.0 to 782.6~\cm{} (12.77 to 14.27~$\mu$m). 
These observations were conducted while SOFIA flew at an altitude of 13.1~km. 
For both settings, the slit length was $\simeq 2\arcsec$ long and we chose the 2\farcs4 wide slit.
An additional High--Low setting centered around 1335~\cm{} (7.5~$\mu$m) will be discussed elsewhere.

All EXES data were reduced using the Redux pipeline \citep{clarke_2015}.
The median spectral resolving power, $R=\lambda/\Delta\lambda$, was empirically determined from telluric ozone (O$_3$) lines to be about 70,000 for both settings.
The resulting spectral resolution is thus $\simeq 4.3$~\kms.
We calculate the radial velocity of \rleo{} with respect to Earth during the observing flight to be $-22.4$~\kms, which imparts a blueshift to the stellar features by $\simeq 0.055$~\cm{} ($\simeq 0.001~\mu$m) at 740~\cm{} ($\simeq 13.514~\mu$m).

These observations have been complemented with photometric data of the IRAS, WISE, DIRBE, 2MASS, Gaia, and Hipparcos catalogs.
Additional measures acquired with Johnson filters have also been used.
All of them have been taken from VizieR\footnote{\url{https://vizier.u-strasbg.fr/index.gml}}.

\section{Results}
\label{sec:results}

The thinness of the atmosphere at 13~km along with the high excitation of these lines significantly reduced the opacity of the telluric CO$_2$ features and highly increased the atmospheric transmission in the observed spectral range.
Lines from 7 ro-vibrational bands of the main isotopologue are detected ($01^10_1-00^00_1$, $10^00_1-01^10_1$, $02^20_1-01^10_1$, $11^10_1-10^00_1$, $11^10_1-02^20_1$, $20^00_1-11^10_1$, $20^00_2-11^10_2$, and $11^10_2-10^00_2$; Figs.~\ref{fig:f1} and \ref{fig:f2} -- see a brief description of CO$_2$ in appendix~\ref{sec:co2}).
Lines of the hot bands $10^00_1-01^10_1$ and $11^10_1-02^20_1$ of $^{13}$CO$_2$ have also been detected (not included in Figs.~\ref{fig:f1} and \ref{fig:f2} for the sake of clarity).
No line of $^{17}$OCO and $^{18}$OCO has been found, as expected given the low abundance of the oxygen isotopes \citep[$^{12}$C/$^{13}$C~$\simeq 10$, $^{16}$O/$^{17}$O~$\simeq 450$, and $^{16}$O/$^{18}$O~$\simeq 550$;][]{hinkle_2016}.

The observed lines are seen only in emission with a maximum intensity of 10\% above the continuum.
They are found in the blue-shifted wings of the telluric CO$_2$ features, which are usually much stronger and always in absorption.
The same settings were used during the observation of other sources in the same observing campaign and no similar feature stood out.
Hence, they are not instrumental artifacts but real lines that were separated from their telluric counterparts thanks to a high Doppler shift during the observing run.

The low-$J$ lines of the fundamental band are blocked by the atmosphere but a few lines with $J\gtrsim 70$ can be identified in the spectrum.
The detected lines of the hot and combination bands, for which the Earth atmosphere is more transparent, involve ro-vibrational levels with $J\ge 5$.
The highest excitation ro-vibrational level involved in the detected lines is at $\simeq 7000$~K.
Additional emission lines have been detected in the spectrum but not identified.
They are probably produced by even higher excitation bands of CO$_2$ (e.g., $03^30_1-02^20_1$, $12^20_1-03^30_1$, $31^10_1-22^20_1$, $31^10_2-22^20_2$, or $20^01_2-11^11_2$) or $^{13}$CO$_2$ that cannot be unmistakably identified.

The CO$_2$ emission lines are not systematically accompanied by noticeable blue-shifted absorption features above the detection limit.
It probably occurs because the population of the upper vibrational states of each detected band is very high and the emission covers the absorption component.
This effect could be a consequence of the strong radiation field emitted by \rleo{} and the dusty component of the envelope (see Sect.~\ref{sec:discussion}).

\begin{figure}
  \centering
  \includegraphics[width=0.475\textwidth]{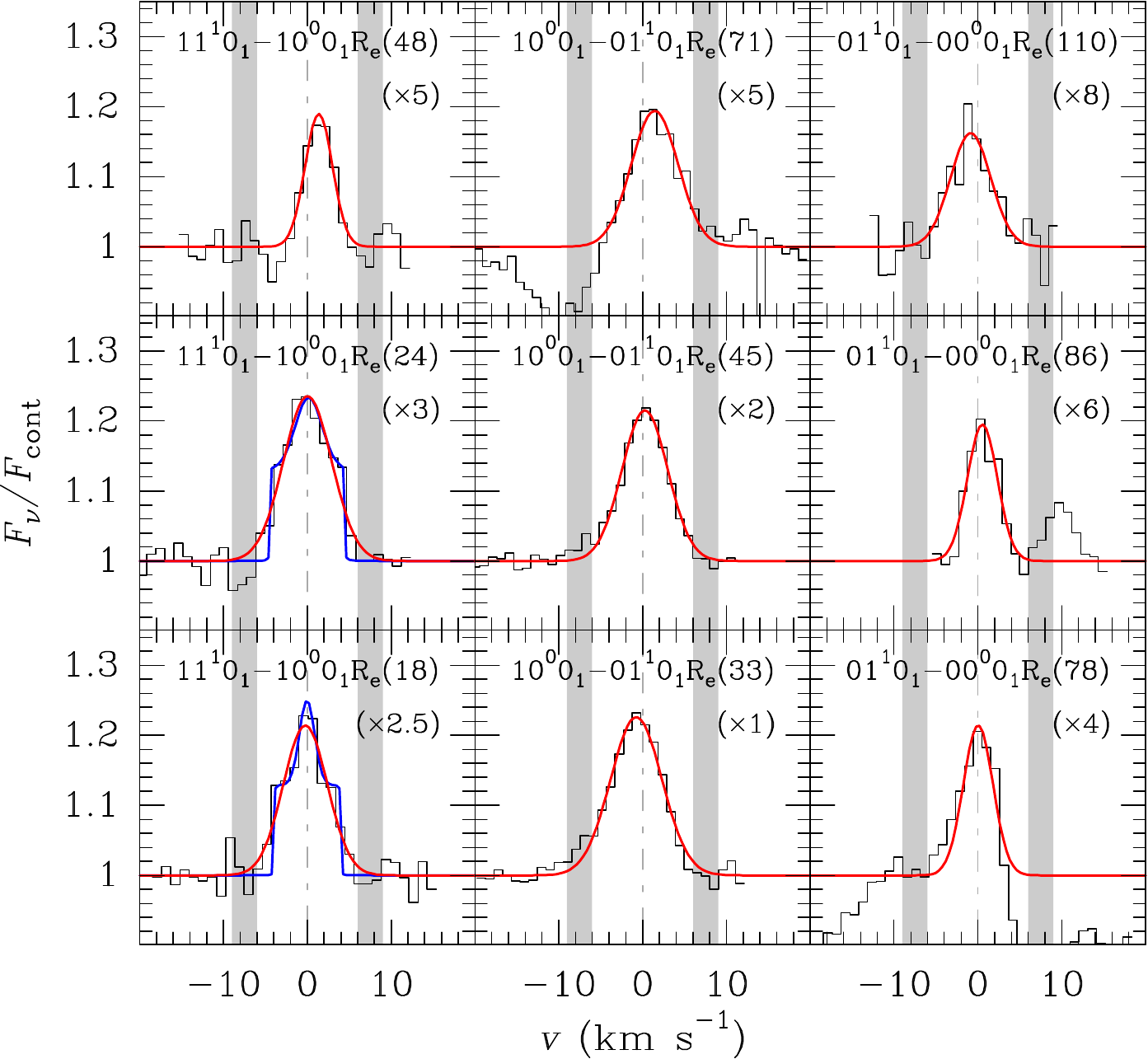}
  \caption{Lines of the R$_e$ branches of bands $11^10_1-10^00_1$, $10^00_1-01^10_1$, and $01^10_1-00^00_1$.
    The baseline has been removed.
    The gray, vertical thick lines represent the terminal gas velocity ranging from 6 to 9~\kms.
    The spectral resolution is $\simeq 4.3$~\kms.
    The red fits have been done with Gaussians.
    The blue fits consider a Gaussian and a rectangular function.}
  \label{fig:f3}
\end{figure}

The comparison of different lines of bands $11^10_1-10^00_1$, $10^00_1-01^10_1$, and $01^10_1-00^00_1$ (Fig.~\ref{fig:f3}) suggests that there is not a significant velocity difference between them.
They are roughly single-peaked lines approximately centered at the systemic velocity and totally delimited by the terminal gas expansion velocity.
The average FWHM of the emission lines is $\simeq 5$~\kms.
Considering that the spectral resolution of our observations is $\simeq 4.3$~\kms, the average deconvolved FWHM is $\simeq 2.5$~\kms{} but the broadest lines show a deconvolved FWHM of $\simeq 5$~\kms.
Therefore, the lines are formed at the beginning of the acceleration region, where the gas expands at velocities up to $\simeq 2.5$~\kms.
The low-$J$ lines of band $11^10_1-10^00_1$ seem to comprise two components: a very narrow peak centered at the systemic velocity and a flat-topped contribution, which probably comes from already accelerated gas.

\begin{figure*}
  \centering
  \includegraphics[width=\textwidth]{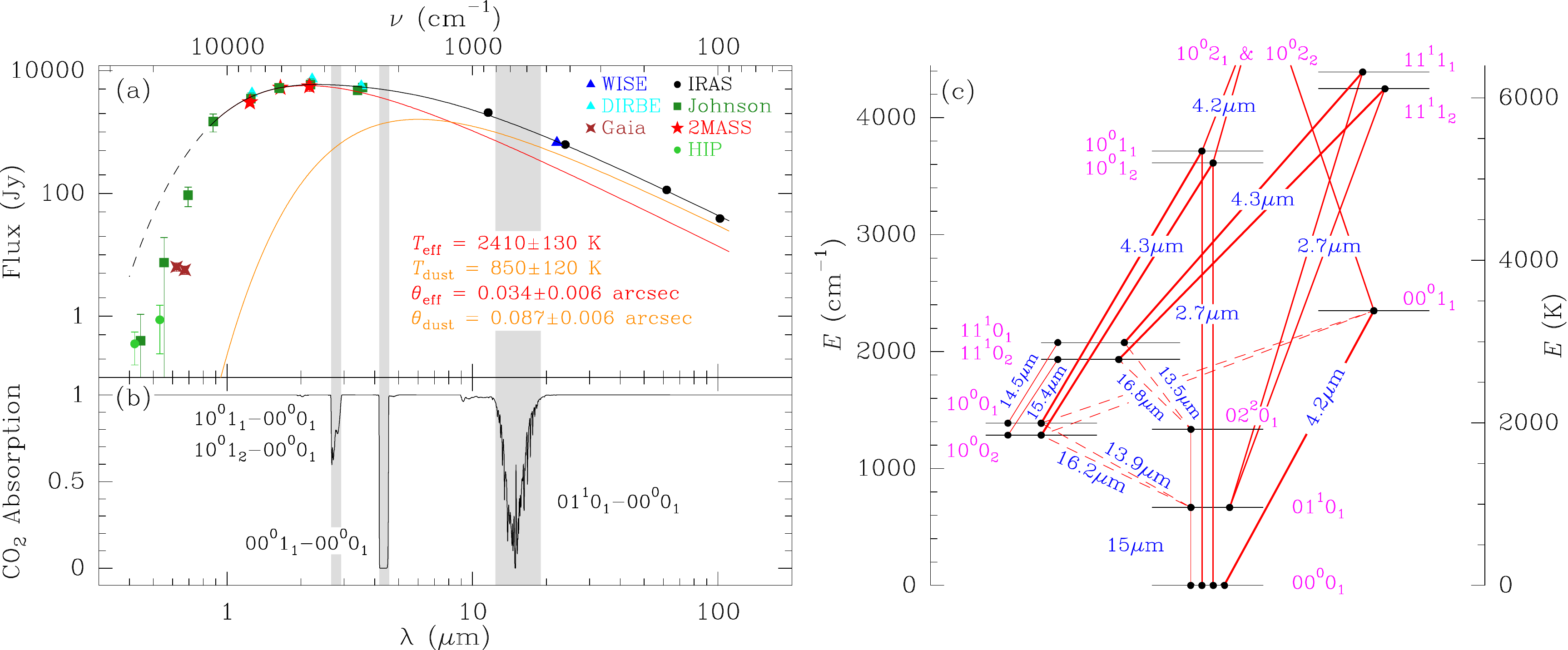}
  \caption{(a) Continuum emission of \rleo{} based on photometric observations available in several databases.
    This data set has been fitted with two black-bodies between 0.8 and 100~$\mu$m (solid black curve) to represent the star (``eff'') and the dusty component of the envelope (``dust'').
    The dashed black curve is an extrapolation of this fit below 0.8~$\mu$m.
    $T_X$ and $\theta_X$ are the temperature and the angular diameter of the $X$ emitting region.
    The vertical gray bands indicate the spectral ranges covered by the strongest CO$_2$ bands.
    Neither dust opacity nor scattering have been considered in our fits.
    (b) Low spectral resolution absorption spectrum of CO$_2$ calculated under LTE at a temperature of 1500~K and with a column density of $5\times 10^{18}$~\cmm.
    The strongest bands are also highlighted in gray.
    (c) Vibrational energy diagram of CO$_2$ with the most important vibrational states involved in the observed bands.
    The possible vibrational transitions have been plotted as red lines with increasing thickness as their $A$-Einstein coefficient grows (dashed lines for the lowest coefficients).}
  \label{fig:f4}
\end{figure*}

We have analyzed the strongest CO$_2$ bands with a ro-vibrational diagram (appendix~\ref{sec:rovib.diagram}).
The continuum intensity has been estimated by fitting the photometric data beyond 0.7~$\mu$m (Fig.~\ref{fig:f4}a).
The mass-loss rate of \rleo{} is relatively low \citep[$\simeq 1.0\times 10^{-7}$~\mlr; e.g.,][]{ramstedt_2014}, hence the dust grain density is also low.
Nevertheless, they contribute significantly to the continuum emission, which can be described to first approximation as two black-bodies: a compact one at $\simeq 2400$~K mainly associated to the central star and a more extended black-body at a temperature of $\simeq 850$~K.
The warm black-body would represent the bulk of dust emission, which comes from a region with a diameter of $\simeq 0\farcs09$.
This value is a lower limit since dust emission would be better described by a more extended gray-body in a more detailed model.

\begin{figure}
  \centering
  \includegraphics[width=0.475\textwidth]{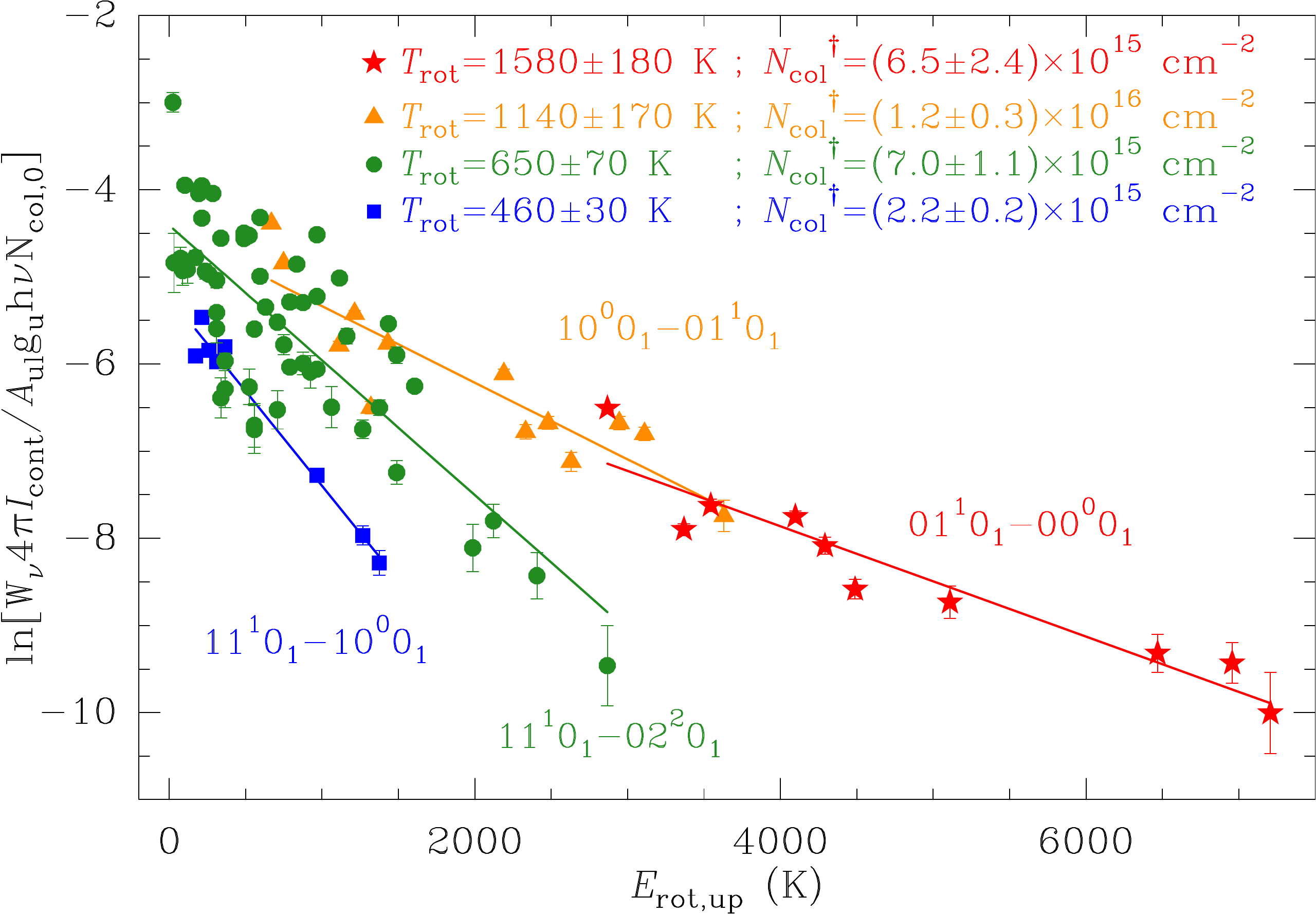}
  \caption{Ro-vibrational diagram of the strongest observed CO$_2$ bands.
    The chosen stellar effective temperature is 2750~K.
    We defined $N_\subscript{col}^\dagger=N_\subscript{col,vib}\left(\theta_\subscript{em}^2/\theta_\subscript{cont}^2\right)\left[(\theta_b^2+\theta_\subscript{cont}^2)/(\theta_b^2+\theta_\subscript{em}^2)\right]$, where $N_\subscript{col,vib}$ is the column density of molecules in the upper vibrational state under study.}
  \label{fig:f5}
\end{figure}

The CO$_2$ lines can be grouped into three different populations, namely warm, hot, and very hot, with approximate temperature populations of 550, 1150, and 1600~K.
The dispersion of the data set related to the $11^10_1-02^20_1$ band is high and the uncertainty of the rotational temperature derived from the fit could be underestimated.
Assuming only a warm population is the most conservative approach.
Interestingly, the bands involving higher energy vibrational states ($11^10_1-10^00_1$ and $11^10_1-02^20_1$) show lower rotational temperatures (Sect.~\ref{sec:discussion}).

The emitting region related to each population can be roughly determined in the envelope.
The lack of a pure rotational spectrum for CO$_2$ prevents an efficient radiative rotational relaxation and implies that CO$_2$ is rotationally under LTE throughout the whole envelope.
The kinetic temperature for the CSE of \rleo{} is not well known but we can assume it to follow the power-law $\propto r^{-\alpha}$, where $\alpha\simeq 0.7$.
This exponent is commonly used to analyze the emission of O-rich stars \citep[e.g.,][]{decin_2006,velilla-prieto_2015,sanchez-contreras_2015} as well as being very similar to that derived from observations of the outer envelope of the carbon rich star \irc{} \citep[$\simeq 0.68$;][]{guelin_2018}, where the gas should be as rarefied as in the envelope of \rleo.
Therefore, the CO$_2$ populations at 1600, 1150, and 550~K are located at 2.2, 3.5, and 10\rstar{} from the center of \rleo, respectively.
Considering that the gas in these three regions is being accelerated, kinematic effects such as line broadening are expected as the bulk of the line emission comes from higher radii.
The dual component detected in the low-$J$ lines of band $11^10_1-10^00_1$ could be evidence of these effects (Fig.~\ref{fig:f3}).

Considering an angular stellar diameter of 0\farcs028, the angular sizes of the emitting regions are $\simeq 0\farcs06$, 0\farcs1, and 0\farcs28.
The corresponding ratios of beam filling factors are 0.49, 1.4, and 10.5, respectively.
Hence, the resulting column densities for the populations in the vibrational states $01^10_1$, $10^00_1$, and $11^10_1$ described by the ro-vibrational diagram are $(1.3\pm 0.5)\times 10^{16}$, $(8.6\pm 2.4)\times 10^{15}$, and $(4.4\pm 0.7)\times 10^{14}$~\cmm, respectively.
These partial results allows us to establish only a lower limit for the total column density in the envelope: $N_\subscript{col}\gtrsim (2.2\pm 0.8)\times 10^{16}$~\cmm.
A larger continuum source would increase this lower limit.

\section{Discussion and final remarks}
\label{sec:discussion}

The continuum emission peaks around 2.5~$\mu$m (Fig.~\ref{fig:f4}a) and it is capable of exciting a significant fraction of the CO$_2$ molecules up to high energy vibrational states thanks to the strong CO$_2$ bands $10^01_1-00^00_1$ and $10^01_2-00^00_1$, which are at $\simeq 2.7~\mu$m, and $00^01_1-00^00_1$, at $\simeq 4.2~\mu$m (Fig.~\ref{fig:f4}b).
The stellar radiation field dominates the excitation of CO$_2$ up to a few stellar radii beyond the photosphere.
States $10^01_1$ and $10^01_2$ serve as a bridge to efficiently populate states $10^00_1$ and $10^00_2$ from the ground strengthening the emission of bands $10^00_1-01^10_1$ and $10^00_2-01^10_1$ (Fig.~\ref{fig:f4}c).
On the contrary, states $11^10_1$ and $11^10_2$ are less efficiently populated from the vibrational ground state but they are from $01^10_1$ via $11^11_1$ and $11^11_2$.
States $11^10_1$ and $11^10_2$ are thus expected to become significantly populated preferentially at several stellar radii from the star due to the increase of the continuum flux at 14~$\mu$m coming from dust as the stellar emission weakens.
This scenario could explain why the rotational temperatures of bands $11^10_1-02^20_1$ and $11^10_1-10^00_1$ are lower than that of $10^00_1-01^10_1$, as the bulk emission of each band comes from different regions of the envelope, where the upper states are more efficiently excited.

Many AGB and semi-regular (SR) stars have shown CO$_2$ features in \textit{ISO} and \textit{Spitzer} observations (Sect.~\ref{sec:introduction}) but only the CO$_2$ column densities and abundances of a handful of them have been estimated.
The derived column densities are between $2.0\times 10^{16}$~\cmm{} for R~Cas \citep{markwick_2000} and $1.0\times 10^{19}$~\cmm{} for EP~Aqr \citep{cami_2000}.
Hence, a column density of $\simeq 5\times 10^{17}$~\cmm{} can be adopted as typical for AGBs and SRs with a high dispersion of a factor of 20.
The column density for the vibrationally excited CO$_2$ that we have derived for \rleo{} ($\simeq 2\times 10^{16}$~\cmm) indicates that the amount of CO$_2$ molecules distributed among other vibrational states could be as high as 80\%.

We have done a simple model of \rleo{} to estimate the total column density of CO$_2$ with the code developed by \citet{fonfria_2008} to model the molecular emission of AGB stars.
We have adopted a stellar temperature of 2750~K, a mass-loss rate of $1.0\times 10^{-7}$~\mlr, a solar He abundance with respect to H$_2$ \citep[$\simeq 0.17$;][]{asplund_2009}, a kinetic temperature of $2750(1\rstar/r)^{0.7}$~K, CO$_2$ under rotational LTE, gas accelerated from the photosphere up to 3\rstar{} and expanding at 8~\kms{} beyond, and a thin dusty component outwards from 3\rstar.
A vibrational temperature profile of $\simeq 2750(1\rstar/r)^{0.9}$~K, which is steeper than the kinetic temperature, is needed to produce lines of the $01^10_1-00^00_1$ band with no absorption component in the observed range.
The maximum vibrational temperatures for higher energy states can be as high as 5000~K.
The adoption of temperatures similar to that derived from the $01^10_1-00^00_1$ band results in lines of bands $10^00_1-01^10_1$, $11^10_1-10^00_1$, or $11^10_1-02^20_1$ incompatible with the observations.
With these parameters, a good agreement between the observations and the model is achieved by chosing an abundance with respect to H$_2$ of $2.5\times 10^{-5}$.
The resulting total CO$_2$ column density is $\simeq 7\times 10^{17}$~\cmm{} with a conservative uncertainty of a factor of 3.
These results involve \textit{all} the vibrational states, which explains the different values derived from the ro-vibrational diagram.
Most of the CO$_2$ molecules are in unobserved vibrational states.
In particular, about 10\% of them are in the ground and thus impossible to observe without a space-based facility.
The synthetic spectrum can be seen in Figs.~\ref{fig:f1} and \ref{fig:f2}.

This result is similar to the column density found for R~Tri \citep[$\simeq 5.2\times 10^{17}$~\cmm;][]{baylis-aguirre_2020}, which has the same mass-loss rate ($\simeq 1.1\times 10^{-7}$~\mlr) as \rleo.
The CO$_2$ column density for $o$~Cet, which has a mass-loss rate twice as high as \rleo, is $\simeq 2\times 10^{17}$~\cmm{} and it is $1.0\times 10^{19}$~\cmm{} for EP~Aqr, which has a mass-loss rate of $2.3-3.1\times 10^{-7}$~\mlr{} \citep{knapp_1998,debeck_2010}.
This discrepancy suggests that either the CO$_2$ column density does not depend strongly on the mass-loss rate or it has not been accurately estimated.
Thus, a systematic study of the CO$_2$ emission in O-rich is needed to understand how this molecule forms.

The detection of the CO$_2$ bands presented in the current work demonstrates the incomparable ability of SOFIA/EXES to observe the infrared spectrum of molecules in space, even when they are highly perturbed by the Earth atmosphere, due to the combination of (1) high spectral resolution, (2) the very thin atmosphere that exists above the stratosphere, and (3) a favorable observation date to have the best Doppler shift with respect to the telluric absorption, which is crucial to get reliable line profiles after the removal of the baseline.
Despite the CO$_2$ column density presented in this work was derived simplistically with substantial uncertainties, a careful analysis that considers CO$_2$ under rotational LTE can improve the accuracy of the results.

\begin{acknowledgements}

We thank the anonymous referee for his/her valuable comments about the manuscript.
The research leading to these results has received funding support from the European Research Council under the European Union's Seventh Framework Program (FP/2007-2013) / ERC Grant Agreement n. 610256 NANOCOSMOS.
EJM acknowledges financial support for this work through award \#06\_0144 which was issued by USRA and provided by NASA.
MJR and EXES observations are supported by NASA cooperative agreement 80NSSC19K1701.
Based on observations made with the NASA/DLR Stratospheric Observatory for Infrared Astronomy (SOFIA).
SOFIA is jointly operated by the Universities Space Research Association, Inc. (USRA), under NASA contract NAS2-97001, and the Deutsches SOFIA Institut (DSI) under DLR contract 50 OK 0901 to the University of Stuttgart.

\end{acknowledgements}

\begin{appendix}

\section{Brief description of CO$_2$}
\label{sec:co2}

CO$_2$ is a linear triatomic molecule that belongs to the $D_{\infty h}$ group and, hence, displays no permanent dipole moment.
Its atoms vibrate according to four normal modes, two of which are degenerate.
Two of these modes, $\nu_1(\sigma_g^+)$ and $\nu_3(\sigma_u^+)$, are stretching modes.
The other two degenerate modes are combined into the bending mode $\nu_2(\pi_u)$, which induces a vibrational angular splitting in the ro-vibrational levels with different $e-f$ parity \citep{brown_1975}.
The fundamental band associated to the $\nu_1(\sigma_g^+)$ mode is infrared inactive due to the $g-u$ symmetry with respect to the molecular middle plane.
The absence of non-zero nuclear spins forbids the rotational levels with even or odd $J$, depending on the symmetry of the vibrational state.

The similarity between the energies of the vibrational states $\nu_1(\sigma_g^+)$ and $2\nu_2(\sigma_g^+)$ and other overtones produces a Fermi resonance that results in groups of significantly perturbed states.
For simplicity, we have adopted in this work the notation $v_1v_2^lv_{3,r}$ for the vibrational states, where $v_i$ is the $i$-th vibrational number, $l$ the vibrational angular momentum number related to the bending mode $\nu_2$ and the index $r=1,2,\ldots$ orders the vibrational levels of each Fermi resonant group in decreasing order of energy \citep{rothman_1981}.
The frequencies of the lines have been taken from the HITRAN Database \citep{gordon_2017}.

\section{Additional observations}

\begin{figure*}
  \centering
  \includegraphics[width=0.98\textwidth]{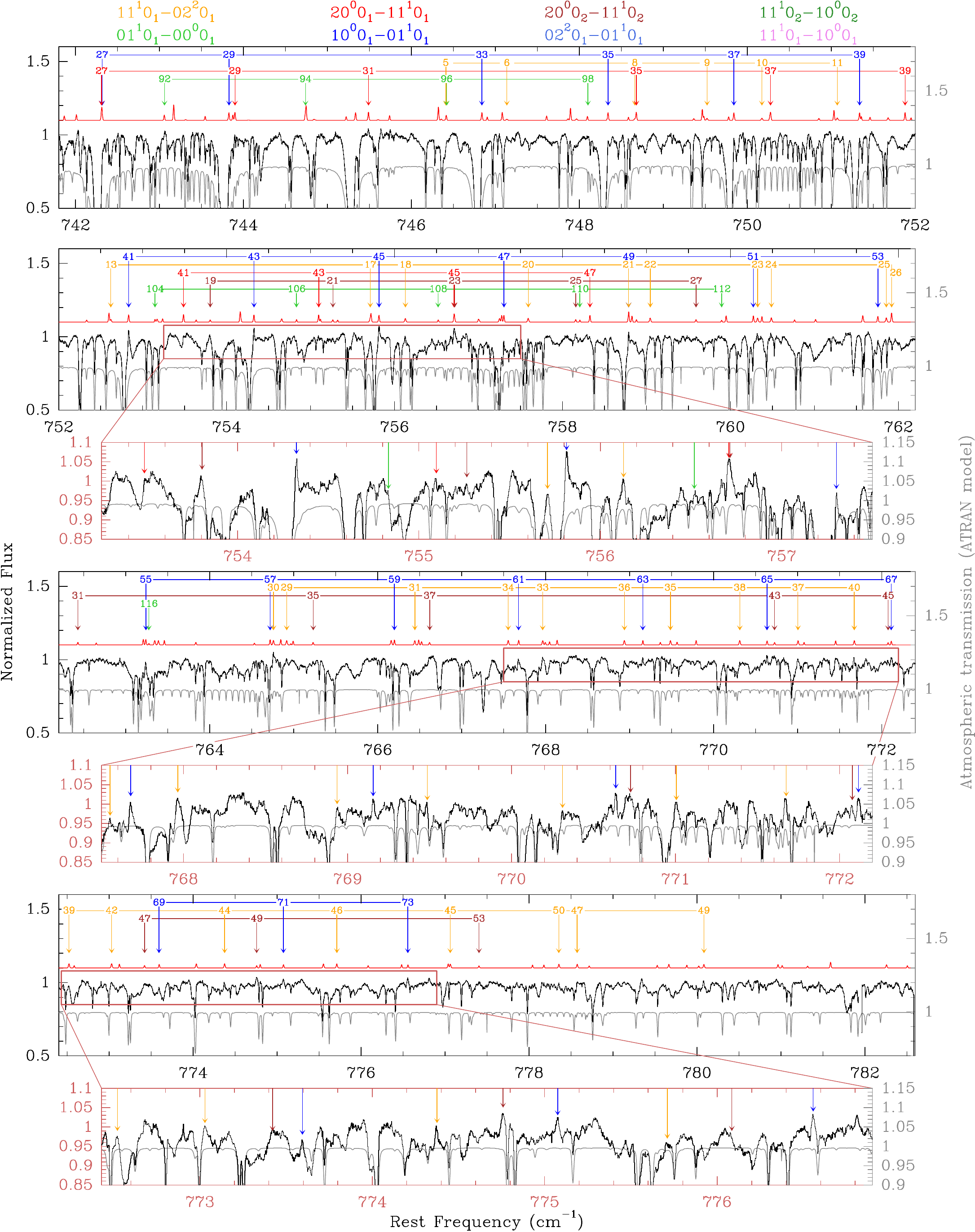}
  \caption{Spectrum of \rleo{} in the spectral range $742-782$~\cm.
    See caption of Fig.~\ref{fig:f1}.}
  \label{fig:f2}
\end{figure*}

The observations covered a total spectral range from 701.0 to 782.6~\cm.
  Fig.~\ref{fig:f1} shows the range $701.0-741.8$~\cm, which includes most of the strongest detected lines.
  The rest of the total spectral range, $741.8-782.6$~\cm, contains a lower number of strong features but there are still a significant amount of weaker CO$_2$ lines (Fig.~\ref{fig:f2}).

\section{The ro-vibrational diagram}
\label{sec:rovib.diagram}

We have made a ro-vibrational diagram for the strongest observed emission lines by using the well-known formula by \citet{goldsmith_1999} adapted to a normalized infrared spectrum:
\begin{equation}
  \label{eq:rovib.diagram} \ln\left[\frac{W 4\pi I_{\nu,\subscript{cont}}}{A_{ul}g_uh\nu N_\subscript{col,0}}\right]\simeq
  y_0-\frac{hcE_\subscript{rot,up}}{k_\subscript{B}T_\subscript{rot}},
\end{equation}
where $W$ is the integral of the observed line over the frequency (\cm), $\nu$ is the rest frequency (\cm), $E_\subscript{rot,up}$ the rotational energy of the upper level involved in the transition, $T_\subscript{rot}$ the rotational temperature of the upper vibrational state, and $I_{\nu,\subscript{cont}}$ is the continuum emission measured from Earth.
$N_{\subscript{col},0}$ is a column density set arbitrarily to $1.0\times 10^{15}$~\cmm{} to get dimensionless arguments for the logarithms.
We define the quantity $y_0$ as:
\begin{equation}
  \label{eq:y-intercept}
  y_0=\ln\left[\frac{N_\subscript{col,vib}}{N_\subscript{col,0}Z_\subscript{rot}}\left(\frac{\theta_\subscript{em}^2}{\theta_\subscript{cont}^2}\frac{\theta_b^2+\theta_\subscript{cont}^2}{\theta_b^2+\theta_\subscript{em}^2}\right)\right],
\end{equation}
where $N_\subscript{col,vib}$ is the column density  of the upper vibrational state involved in the band, $Z_\subscript{rot}$ the rotational partition function, and $\left(\theta_\subscript{em}^2/\theta_\subscript{cont}^2\right)\left[(\theta_b^2+\theta_\subscript{cont}^2)/(\theta_b^2+\theta_\subscript{em}^2)\right]$ the ratio of the beam filling factors for the line emission and continuum sources.
$\theta_\subscript{em}$ and $\theta_\subscript{cont}$ are the diameters of the line and the continuum emitting regions.
$\theta_\subscript{b}$ is the angular size of the convolution of the Point Spread Function (PSF), the atmospheric seeing, and any possible instrumental effect.
In this work, we have assumed that $\theta_b\simeq 3\arcsec$.

\end{appendix}

\end{document}

%% file: definitions.tex
\newcommand{\subscript}[1]{\textnormal{\tiny{#1}}}

\newcommand{\rstar}{\ensuremath{R_\star}}
\newcommand{\kms}{km~s$^{-1}$}
\newcommand{\cm}{cm$^{-1}$}
\newcommand{\cmm}{cm$^{-2}$}

\newcommand{\mlr}{M$_\odot$~yr$^{-1}$}

\newcommand{\irc}{IRC+10216}
\newcommand{\rleo}{R~Leo}

\newcommand{\sci}{Sci}

\newcommand{\jms}{JMoSp}

\newcommand{\jai}{JAI}


%% file: paper.bbl
\begin{thebibliography}{}
\bibitem[Ag\'undez et al.(2008)]{agundez_2008}                   Ag\'undez, M., Fonfr\'ia, J. P., Cernicharo, J., Pardo, J. R. \& Gu\'elin, M., 2008, \aap, 479, 493
\bibitem[Ag\'undez et al.(2014)]{agundez_2014}                   Ag\'undez, M., Cernicharo, J., Decin, L., Encrenaz, P. \& Teyssier, D., 2014, \apj, 790, L27
\bibitem[Ag\'undez et al.(2020)]{agundez_2020}                   Ag\'undez, M., Mart\'inez, J. I., de Andr\'es, P. L., Cernicharo, J. \& Mart\'in-Gago, J. A., \aap, 637, A59
\bibitem[Asplund et al.(2009)]{asplund_2009}                     Asplund, M., Grevesse, N., Sauval, A. Jacques \& Scott, P., 2009, \araa, 47, 481
\bibitem[Baylis-Aguirre et al.(2020)]{baylis-aguirre_2020}       Baylis-Aguirre, D. K., Creech-Eakman, M. J. \& G\"uth, T., 2020, \mnras, 493, 807
\bibitem[Bernath \& Hinkle(1989)]{bernath_1989}                  Bernath, P. F., Hinkle, K. H. \& Keady, J. J., 1989, \sci, 244, 562
\bibitem[Betz(1981)]{betz_1981}                                  Betz, A. L., 1981, \apj, 244, L103
\bibitem[Bieging et al.(2000)]{bieging_2000}                     Bieging, J. H., Shaked, S. \& Gensheimer, P. D., 2000, \apj, 543, 897
\bibitem[Brown et al.(1975)]{brown_1975}                         Brown, J. M., Hougen, J. T., Huber, K.-P., et al., 1975, \jms, 55, 500
\bibitem[Bujarrabal et al.(1994)]{bujarrabal_1994}               Bujarrabal, V., Fuente, A. \& Omont, A., 1994, \aap, 285, 247
\bibitem[Cami et al.(2000)]{cami_2000}                           Cami, J., Yamamura, I., de Jong, et al., 2000, \aap, 360, 562
\bibitem[Cernicharo et al.(2000)]{cernicharo_2000}               Cernicharo, J., Gu\'elin, M. \& Kahane, C., 2000, \aaps, 142, 181
\bibitem[Cherchneff(2006)]{cherchneff_2006}                      Cherchneff, I., 2006, \aap, 456, 1001
\bibitem[Clarke et al.(2015)]{clarke_2015}                       Clarke, M., Vacca, W.~D., \& Shuping, R.~Y.\ 2015, Astronomical Data Analysis Software an Systems XXIV (ADASS XXIV), ed. A. R. Taylor and E. Rosolowsky. San Francisco: Astronomical Society of the Pacific, 2015., p.~355
\bibitem[De Beck et al.(2010)]{debeck_2010}                      De Beck, E., Decin, L., de Koter, A., et al., 2010, \aap, 523, A18
\bibitem[Decin et al.(2006)]{decin_2006}                         Decin, L., Hony, S., de Koter, A., et al., 2006, \aap, 456, 549
\bibitem[Etoka \& Le Squeren(1997)]{etoka_1997}                  Etoka, S. \& Le Squeren, A. M., 1997, \aap, 321, 877
\bibitem[Fedele et al.(2005)]{fedele_2005}                       Fedele, D., Wittkowski, M., Paresce, F., et al., 2005, \aap, 431, 1019
\bibitem[Fonfr\'ia et al.(2008)]{fonfria_2008}                   Fonfr\'ia, J. P., Cernicharo, J., Richter, M. J. \& Lacy, J. H., 2008, \apj, 673, 445
\bibitem[Goldhaber \& Betz(1984)]{goldhaber_1984}                Goldhaber, D. M. \& Betz, A. L., 1984, \apj, 279, L55
\bibitem[Goldsmith \& Langer(1999)]{goldsmith_1999}              Goldsmith, P. F. \& Langer, W. D., 1999, \apj, 517, 209
\bibitem[Gonz\'alez-Delgado et al.(2003)]{gonzalez-delgado_2003} Gonz\'alez-Delgado, D., Olofsson, H., Kerschbaum, F., et al., 2003, \aap, 411, 123
\bibitem[Gordon et al.(2017)]{gordon_2017}                       Gordon, I. E., Rothman, L. S., Hill, C., et al., 2017, \jqsrt, 203, 3
\bibitem[Gu\'elin et al.(2018)]{guelin_2018}                     Gu\'elin, M., Patel, N., Bremer, M., et al., 2018, \aap, 610, A4
\bibitem[Hinkle \& Barnes(1979)]{hinkle_1979}                    Hinkle, K. H. \& Barnes, T. G., 1979, \apj, 227, 923
\bibitem[Hinkle et al.(2016)]{hinkle_2016}                       Hinkle, K. H., Lebzelter, T. \& Strainero, O., 2016, \apj, 825, 38
\bibitem[Justtanont et al.(1996)]{justtanont_1996}               Justtanont, K., de Jong, T., Helmich, F. P., et al., 1996, \aap, 315, L217
\bibitem[Justtanont et al.(1998)]{justtanont_1998}               Justtanont, K., Feuchtgruber, H., de Jong, T., et al., 1998, \aap, 330, L17
\bibitem[Knapp et al.(1998)]{knapp_1998}                         Knapp, G. R., Young, K., Lee, E., et al., 1998, \apjs, 117, 209
\bibitem[Lord(1992)]{lord_1992}                                  Lord S. D., 1992, NASA Technical Memorandum 103957
\bibitem[Markwick \& Millar(2000)]{markwick_2000}                Markwick, A. J. \& Millar, T. J., 2000, \aap, 359, 1162
\bibitem[McGuire(2018)]{mcguire_2018}                            McGuire, B. A., 2018, \apjs, 239, 17
\bibitem[Ohnaka(2004)]{ohnaka_2004}                              Ohnaka, K., 2004, \aap, 424, 1011
\bibitem[Perrin et al.(1999)]{perrin_1999}                       Perrin, G., Coud\'e du Foresto, V., Ridgway, S. T., et al., 1999, \aap, 345, 221
\bibitem[Ramstedt \& Olofsson(2014)]{ramstedt_2014}              Ramstedt, S. \& Olofsson, H., 2014, \aap, 556, A145
\bibitem[Reiter et al.(2015)]{reiter_2015}                       Reiter, M., Marengo, M., Hora, J. L. \& Fazio, G. G., 2015, \mnras, 447, 3909
\bibitem[Richter et al.(2018)]{richter_2018}                     Richter, M.~J., Dewitt, C.~N., McKelvey, M., et al.\ 2018, \jai, 7, 1840013
\bibitem[Ridgway et al.(1976)]{ridgway_1976}                     Ridgway, S. T., Hall, D. N. B., Kleinmann, S. G., Weinberger, D. A. \& Wojslaw, R. S., 1976, \nat, 264, 345
\bibitem[Rothman \& Young(1981)]{rothman_1981}                   Rothman, L.. S. \& Young, L. D. G., 1981, \jqsrt, 25, 505
\bibitem[Ryde et al.(1997)]{ryde_1997}                           Ryde, N., Eriksson, B. Gustafsson, M. Lindqvist \& H. Olofsson, 1997, \apss, 255, 301
\bibitem[S\'anchez Contreras et al.(2015)]{sanchez-contreras_2015} S\'anchez Contreras, C., Velilla Prieto, L., Ag\'undez, M., et al., 2015, \aap, 577, 52
\bibitem[Sch\"oier et al.(2013)]{schoier_2013}                   Sch\"oier, F. L., Ramstedt, S., Olofsson, H., et al., 2013, \aap, 550, A78
\bibitem[Sloan et al.(2010)]{sloan_2010}                         Sloan, G. C., Matsunaga, N., Matsuura, M., et al., 2010, \apj, 719, 1274
\bibitem[Smolders et al.(2012)]{smolders_2012}                   Smolders, K., Neyskens, P., Blommaert, J. A. D. L., et al., 2012, \aap, 540, A72
\bibitem[Temi et al.(2018)]{temi_2018}                           Temi, P., Hoffman, D., Ennico, K. \& Le, J., 2018, \jai, 7, 1840011
\bibitem[Tsuji et al.(1997)]{tsuji_1997}                         Tsuji, T., Ohnaka, K., Aoki, W. \& Yamamura, I., 1997, \aap, 320, L1
\bibitem[Velilla Prieto et al.(2015)]{velilla-prieto_2015}       Velilla Prieto, L., S\'anchez Contreras, C., Cernicharo, J., et al., 2015, \aap, 575, A84
\bibitem[Velilla Prieto et al.(2017)]{velilla-prieto_2017}       Velilla Prieto, L., S\'anchez Contreras, C., Cernicharo, J., et al., 2017, \aap, 597, A25
\bibitem[Wittkowski et al.(2016)]{wittkowski_2016}               Wittkowski, M., Chiavassa, A., Freytag, B., et al., 2016, \aap, 587, A12
\bibitem[Yamamura et al.(1999)]{yamamura_1999}                   Yamamura, I., de Jong, T. \& Cami, J., 1999, \aap, 348, L55
\end{thebibliography}
